\documentclass[a4paper]{article}

\usepackage[dvips]{graphicx}
\usepackage{epstopdf}
\usepackage{INTERSPEECH2021}
\usepackage{amsmath,amssymb}

\usepackage{bm}
\usepackage{mathptmx} 
\usepackage{times} 
\usepackage{algorithm}
\usepackage{algorithmic}
\usepackage{url}
\usepackage{cite}

\usepackage{lineno,hyperref}

\usepackage{color}
\usepackage{float}
\usepackage{tabularx}

\title{StarGAN-VC+ASR: StarGAN-based Non-Parallel Voice Conversion Regularized by Automatic Speech Recognition}
\name{Shoki Sakamoto$^1$,
      Akira Taniguchi$^1$,
      Tadahiro Taniguchi$^1$
      Hirokazu Kameoka$^2$}
\address{
  $^1$College of Information Science and Engineering, Ritsumeikan University, Japan\\
  $^2$NTT Communication Science Laboratories, NTT Corporation, Japan}
\email{\{sakamoto.shoki, a.taniguchi, taniguchi\}@em.ci.ritsumei.ac.jp, kameoka.hirokazu@lab.ntt.co.jp}

\begin{document}

\maketitle
\begin{abstract}
Preserving the linguistic content of input speech is essential during voice conversion (VC). The star generative adversarial network-based VC method (StarGAN-VC) is a recently developed method that allows non-parallel many-to-many VC. Although this method is powerful, it can fail to preserve the linguistic content of input speech when the number of available training samples is extremely small. To overcome this problem, we propose the use of automatic speech recognition to assist model training, to improve StarGAN-VC, especially in low-resource scenarios.
 Experimental results show that using our proposed method, StarGAN-VC can retain more linguistic information than vanilla StarGAN-VC.
\end{abstract}

\section{Introduction}
Non-parallel many-to-many voice conversion (VC) is a powerful and useful framework for building VC systems~\cite{van2017neural,kameoka2018stargan,qian2019autovc}. 
VC is the task of converting the para-linguistic characteristics contained in a given speech signal without changing the linguistic content (transcription)~\cite{stylianou1998continuous}. Conventionally, most VC systems have been developed based on a parallel dataset, consisting of pairs of utterances in the source and target domains reading the same sentences ~\cite{stylianou1998continuous,desai2010spectral,kaneko2017sequence}. 
However, preparing a large-scale parallel dataset is very expensive, and this poses a potential limitation to both the flexibility and performance of VC systems. 
Recently, non-parallel VC methods have been proposed and have attracted much attention~\cite{hsu2016voice,hsu2017voice,kaneko2018cyclegan}. StarGAN-VC~\cite{kameoka2018stargan} is a non-parallel many-to-many VC method based on a variant of the generative adversarial network (GAN)~\cite{goodfellow2014generative} called StarGAN~\cite{isola2017image}.
This method is particularly attractive in that it can generate converted speech signals fast enough to allow real-time implementation and can generate reasonably realistic sounding speech even with limited training samples.


In StarGAN-VC, the use of cycle consistency and identity mapping losses along with adversarial and domain classification losses encourages the generator to preserve the linguistic content of input speech. 
However, this function can still fail when the number of available training utterances is extremely small. In such cases, the linguistic content of converted speech can often be corrupted, resulting in inaudible speech.
Although non-parallel data are easier to prepare than parallel data, they can be expensive, as in applications such as emotional voice conversion~\cite{rizos2020stargan,Moritani-evc}. 
Hence, there is a need for a method that can overcome this problem that can arise under limited data resources.
A possible solution would be to design StarGAN-VC such that it can make explicit use of linguistic information for model training. 

VC systems may benefit from being used in combination with automatic speech recognition (ASR) systems owing to the recent advances in ASR performance.
One successful example involves the use of phonetic posteriorgrams (PPGs)~\cite{sun2016phonetic, kinnunen2017non, Tian2020}.
Similar to the PPG-based approach, the method proposed in this paper improves the ability of StarGAN-VC to preserve the linguistic content of input speech by using ASR to assist model training.
Specifically, we propose using the ASR results applied to each training utterance to construct phoneme-dependent regularization terms for the latent vectors generated from the encoder in StarGAN-VC. 
These penalty terms are derived based on a Gaussian mixture model (GMM); therefore, the latent vectors become more phonetically distinct, resulting in more intelligible speech.
We call the proposed method StarGAN-VC+ASR. 

The main contributions of this study are as follows. We propose a non-parallel many-to-many VC method (StarGAN-VC+ASR) that combines StarGAN-VC and ASR. We experimentally demonstrate that StarGAN-VC+ASR can produce more intelligible speech than the original StarGAN-VC. 

The remainder of this paper is organized as follows. Section 2 briefly introduces StarGAN-VC, which is the basis of the proposed method. Section 3 describes the proposed method, StarGAN-VC+ASR. Section 4 describes an experiment that demonstrates the performance of StarGAN-VC+ASR. Finally, Section 5 concludes the paper.

\section{Preliminaries: StarGAN-VC}
\label{sec:stargan-vc}
StarGAN-VC is based on a model consisting of a generator, $G$, real/fake discriminator, $D$, and domain classifier, $C$. 
Let $o\in\mathbb{R}^{Q\times T}$ be an acoustic feature sequence, where $Q$ is the feature dimension, and $T$ is the length of the sequence. Let $c\in\{1,\ldots,N\}$ be a target domain code, where $N$ is the number of domains.
$G$ is a neural network (NN) that takes an acoustic feature sequence, $o$, in an arbitrary domain and the target domain code, $c$, as inputs and generates an acoustic feature sequence, $\hat{o}'=G(o,c)$. The network is trained such that $G(o,c)$ becomes indistinguishable from the real speech feature sequence in the domain $c$.
This is made possible by using $D$ and $C$ to train $G$, whose roles are to discriminate fake samples, $G(o,c)$, from real samples, $o'$, and identify the class to which $G(o,c)$ is likely to belong.
\begin{figure}[t]
  \begin{center}
    \includegraphics[width=\linewidth]{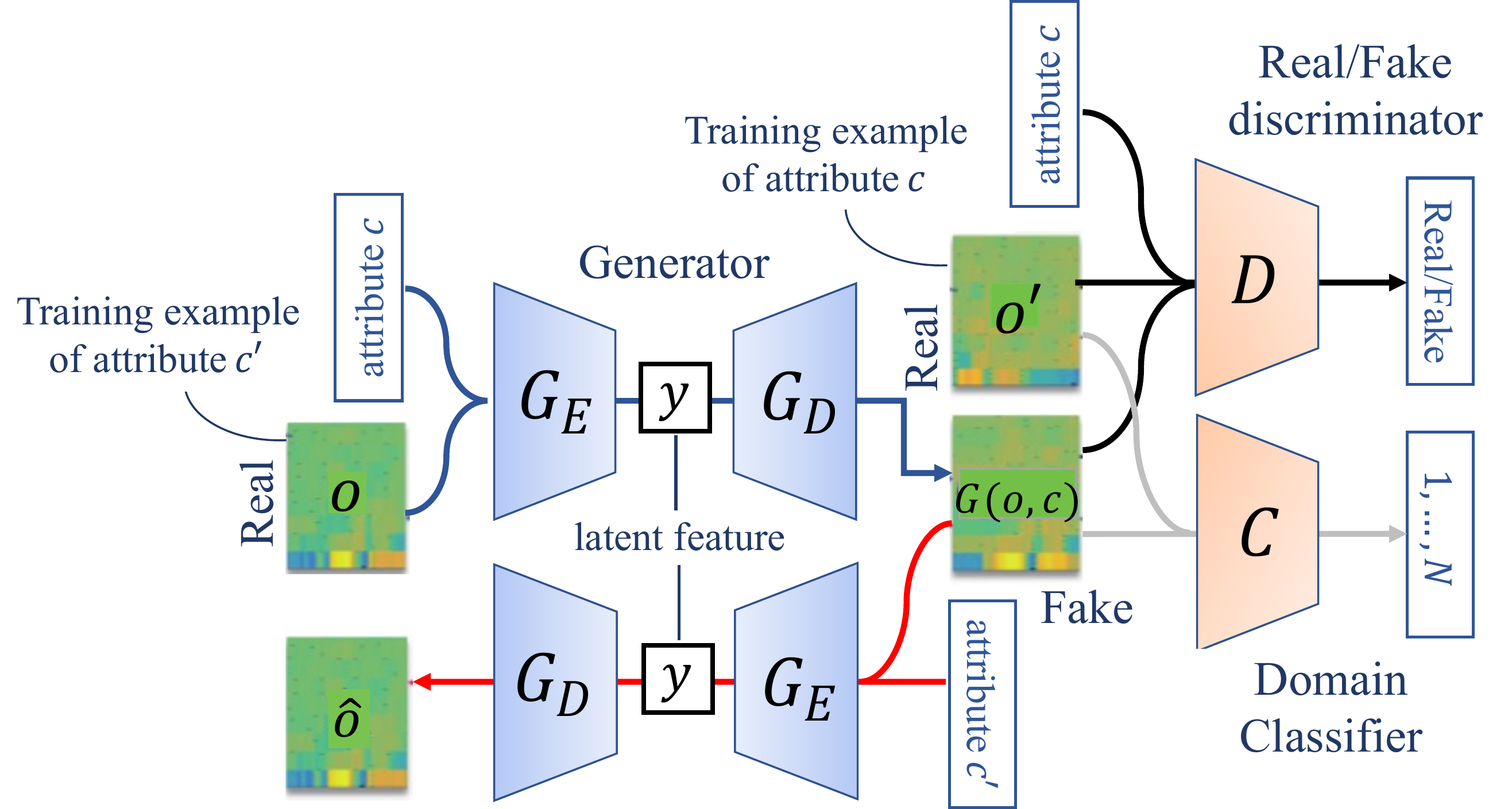}
    \caption{Concept of StarGAN-VC training. $G_{E}$ and $G_{D}$ represent the encoder and decoder in G, respectively.}
    \label{StarGAN_VC_learning}
  \end{center}
\end{figure} 
Figure~\ref{StarGAN_VC_learning} shows the overview of StarGAN-VC training.
$D$ is designed to produce a sequence of probabilities $D(o',c)$, each of which indicates the likelihood that a different segment of input $o'$ is real, whereas $C$ is designed to produce a sequence of class probabilities $p_{C}(c|o')$, each of which indicates the likelihood that a different segment of $o'$ belongs to a particular class. 

For training, StarGAN-VC uses adversarial, cycle consistency, identity mapping, and domain classification losses.

\noindent \textbf{Adversarial\:Loss:}
The adversarial losses are defined for discriminator $D$ and generator $G$ as
\begin{align}
\label{star_adv_loss1}
    \mathcal{L}_{\text{adv}}^D (D) = & - \mathbb{E}_{c\sim p(c), o' \sim p(o' |c)}  \left[\log D(o',c) \right] \nonumber\\ 
    {} & - \mathbb{E}_{o \sim p(o),c\sim p(c)}  \left[\log (1 - D(G(o,c),c)) \right],  \\
\label{star_adv_loss2}
    \mathcal{L}_{\text{adv}}^G (G) = &- \mathbb{E}_{o \sim p(o),c\sim p(c)}  \left[\log D(G(o,c),c) \right],
\end{align}
where $o'\sim p(o'|c)$ denotes a training sample of acoustic feature sequences of real speech belonging to domain $c$, and $o\sim p(o)$ denotes an arbitrary domain.
Equation~\ref{star_adv_loss1} takes a small value when $D$ correctly identifies $G(o,c)$ as fake and $o'$ as real.
By contrast, Equation~\ref{star_adv_loss2} takes a small value when $D$ misclassifies $G(o,c)$ as real. Thus, the subgoals of $D$ and $G$ are to minimize Equations~\ref{star_adv_loss1} and \ref{star_adv_loss2}, respectively.

\noindent \textbf{Domain\:Classification\:Loss:} 
The domain classification losses for classifier $C$ and generator $G$ are defined as
\begin{align}
\label{star_cls_loss1}
\mathcal{L}_{\text{cls}}^C (C) &= - \mathbb{E}_{c\sim p(c), o' \sim p(o' |c)}  \left[\log p_C (c|o') \right], \\
\label{star_cls_loss2}
\mathcal{L}_{\text{cls}}^G (G) &= - \mathbb{E}_{o \sim p(o),c\sim p(c)}  \left[\log p_C (c|G(o,c)) \right].
\end{align}
Equations~\ref{star_cls_loss1} and \ref{star_cls_loss2} have small values when $o'\sim p(o'|c)$ and $G(o,c)$ are correctly classified as belonging to domain $c$ by $C$.
Thus, the subgoals of $C$ and $G$ are to minimize Equations~\ref{star_cls_loss1} and \ref{star_cls_loss2}, respectively.

\noindent \textbf{Cycle-Consistency\:Loss:}
The adversarial and classification losses encourage $G(o,c)$ to become realistic and classifiable, respectively. 
However, using these losses alone does not guarantee that $G$ preserves the linguistic content of the input speech. 
To promote content-preserving conversion, the cycle consistency loss
\begin{align}
\label{star_cyc_loss}
    &\mathcal{L}_{\text{cyc}} (G) \nonumber\\
    &= \mathbb{E}_{c'\sim p(c), o \sim p(o|c'),c\sim p(c)} \left[||G(G(o,c),c') - o ||_\rho \right],
\end{align}
is used, where
$o\sim p(o|c')$ denotes a training sample of acoustic feature sequences of real speech belonging to the source domain $c'$, and $\rho$ is a positive constant.

\noindent \textbf{Identity-Mapping\:Loss:}
To ensure that $G$ keeps its input, $o$, unchanged if $o$ already belongs to the target domain, 
the identity mapping loss is defined as
\fussy
\begin{align}
\label{star_id_loss}
    \mathcal{L}_{\text{id}} (G) = \mathbb{E}_{c'\sim p(c), o \sim p(o|c')}  \left[||G(o,c') - o ||_\rho \right].
\end{align} 

The full loss function is given as
\begin{align}
\label{star_g_loss}
    \mathcal{I}_G (G) = &\mathcal{L}_{\text{adv}}^G (G) + \lambda_{\text{cls}} \mathcal{L}_{\text{cls}}^G (G) \nonumber\\
    &+ \lambda_{\text{cyc}} \mathcal{L}_{\text{cyc}} (G) + \lambda_{\text{id}} \mathcal{L}_{\text{id}} (G), \\
\label{star_d_loss}
    \mathcal{I}_D (D) = & \mathcal{L}_{\text{adv}}^D (D), \\
\label{star_c_loss}
    \mathcal{I}_C (C) = & \mathcal{L}_{\text{cls}}^C (C),
\end{align}
$\lambda_{cls}\geq0$, $\lambda_{cyc}\geq0$, and $\lambda_{id}\geq0$ are regularization parameters that weigh the importance of the domain classification, cycle-consistency, and identity-mapping losses.

\begin{figure}[t]
  \begin{center}
    \includegraphics[width=\linewidth]{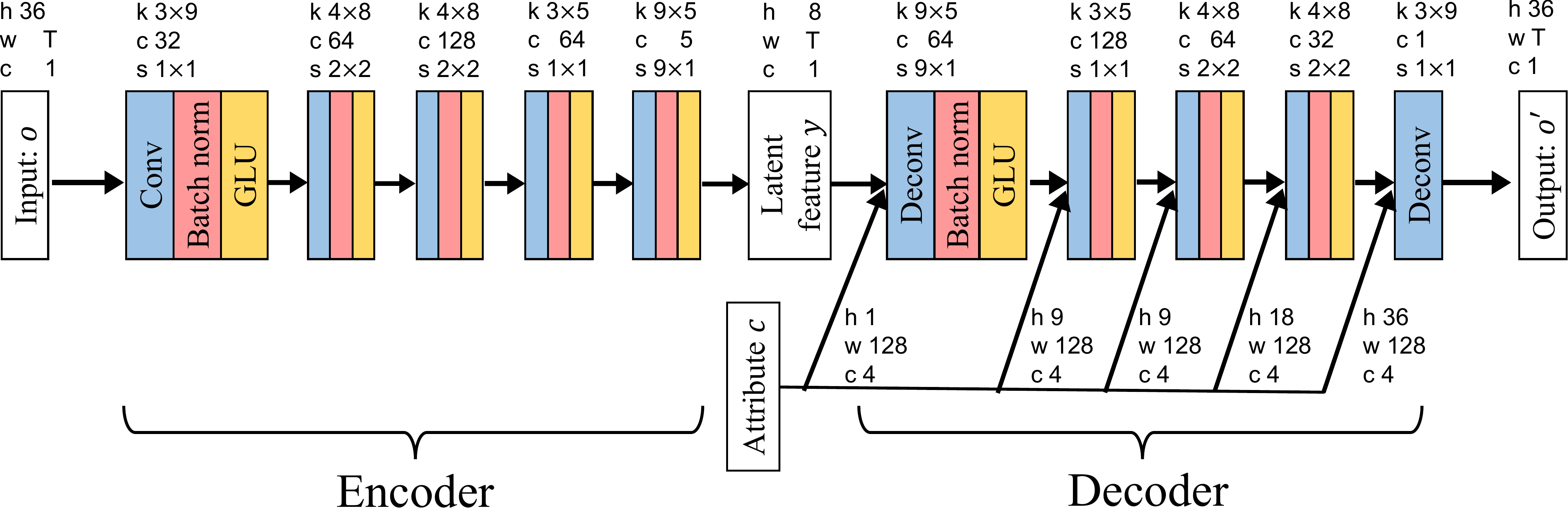}
    \caption{The network architecture of generator $G$.
    In input and latent features, ``h,'' ``w,'' and ``c'' represent height, width, and number of channels, respectively. In addition, ``k,'' ``c, '' and ``s'' denote kernel size, number of channels, and stride size, respectively. ``Conv,'' ``Batch norm,'' ``GLU,'' ``Deconv,'' ``Sigmoid,'' ``Softmax,'' and ``Product'' denote convolution, batch normalization, gated linear unit, transposed convolution, sigmoid, softmax, and product pooling layers, respectively.}
    \label{StarGAN_VC_gen}
  \end{center}
\end{figure}

Regarding network architectures, we assume an encoder-decoder type architecture for $G$, as shown in Figure~\ref{StarGAN_VC_gen}. The encoder is responsible for extracting a speaker-independent latent feature sequence $y$, which is expected to correspond to the linguistic content of the input utterance. The decoder, on the other hand, is responsible for reconstructing an acoustic feature sequence using both target domain codes $c$ and $y$. 

\section{StarGAN-VC+ASR}
\begin{figure*}[t]
  \begin{center}
    \includegraphics[width=0.8\linewidth]{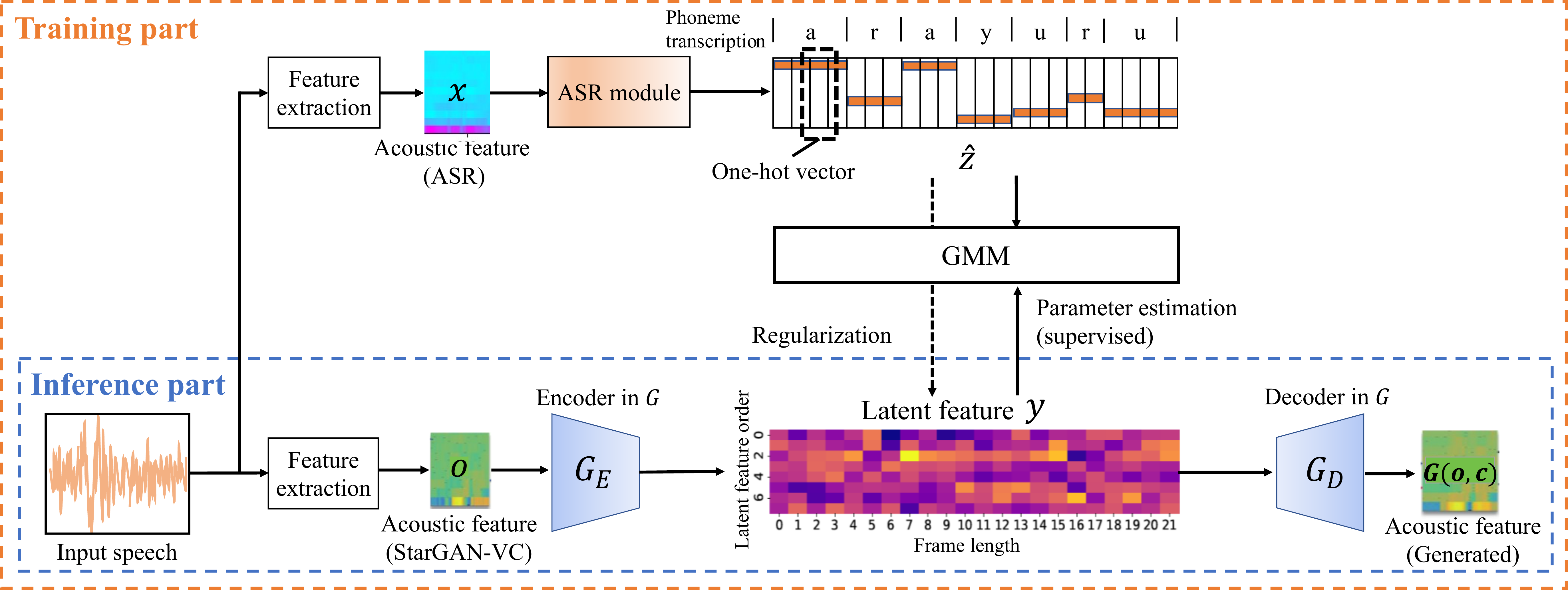}

    \caption{Overview of StarGAN-VC+ASR training}
        \vspace{-5mm}
    \label{asr_overview}
  \end{center}
\end{figure*}

Here, we present the idea of StarGAN-VC+ASR.
Figure~\ref{asr_overview} shows the overview of the StarGAN-VC+ASR training. 

\subsection{GMM-based phoneme model in latent space}
$G$ has an encoder-decoder structure similar to variational autoencoders (VAEs)~\cite{kingma2013auto}.
The encoder part in $G$ is responsible for extracting a speaker-independent latent feature sequence, $y$, corresponding to the linguistic content of the input.
Now, to associate each element of $y$ with the phoneme in the corresponding frame of the input speech, we consider constructing a prior distribution over $y$ based on the result of applying ASR to that speech in advance.
Let $P=\{1,2, \ldots, K\}$ denote a set of phoneme indices, where $K$ denotes the number of phonemes. 
In the ASR step, 
the input is the acoustic feature sequence of $d$th utterance, denoted as
$x^d = x^d_{1:T_d} = (x^d_1,x^d_2, \ldots, x^d_{T_d})$, where $T_d$ and $t$ denote the frame length and index, respectively.  
The ASR output is assumed to be a sequence of phoneme indices $\hat{z}^d=\hat{z}^d_{1:T_d}=\text{ASR}(x^d_{1:T_d})$ of the same length. 
The encoder part in $G$ takes the acoustic feature sequence $o^d$ of the same utterance as the input and produces a latent feature sequence $y^d=y^d_{1:T_d}=(y^d_1,y^d_2, \ldots, y^d_{T_d})$.
To ensure that the latent features corresponding to the same phoneme are close to each other, we assign a single Gaussian distribution to each phoneme index and model a prior distribution of $y^d$ according to $\hat{z}^d$.
Given $y^d$ and $\hat{z}^d$, the maximum likelihood estimates of the mean and covariance of each Gaussian can be obtained as
\begin{align}
\label{mu_p}
    \mu_p =& 
    \frac{1}{\sum^D_{d=1}\#(T_p(\hat{z}^d))}\sum^D_{d=1}\sum_{t \in T_p(\hat{z}^d)} y^d_t , \\
\label{Sigma_p}
    \Sigma_p =&
    \frac{1}{\sum^D_{d=1}\#(T_d(\hat{z}^d))}\sum^D_{d=1}\sum_{t \in T_p(\hat{z}^d)} \left(y^d_t - \mu_p\right)\left(y^d_t - \mu_p\right)^T,
\end{align}
where $p$ and $D$ denote the phoneme index in $P$ and the number of training utterances, respectively. 
$T_p(\hat{z}^d)$ denotes the set consisting of $t$ that satisfies $\hat{z}_t^d = p$, and $\#(\cdot)$ denotes the number of elements in a set.
In the following, we use $\mu_{P}$ and $\Sigma_{P}$ to denote the sets $\mu_P=\{\mu_1,\mu_2, \ldots, \mu_{\#(P)}\}$ and $\Sigma_P=\{\Sigma_1,\Sigma_2, \ldots, \Sigma_{\#(P)}\}$, respectively.

\subsection{Phoneme-dependent regularization loss}

The training process of StarGAN-VC+ASR consists of two sta-ges. 
The first stage corresponds to the original StarGAN-VC training described in Section \ref{sec:stargan-vc}.
In the second stage, $G$ is updated by using $\mu_{P}$ and $\Sigma_{P}$ obtained by Equations \ref{mu_p} and \ref{Sigma_p} for regularization. 
Specifically, we assume 
\begin{align}
    p(y_t^d) = \mathcal{N}\left(y_t^d \mid \mu_{\hat{z}^d_t},\Sigma_{\hat{z}^d_t}\right),
\end{align}
as the prior distribution of $y^d$, 
and define 
\begin{align}
\mathcal{L}_{\text{ASR}} (G) =
- \sum_{d=1}^{D} 
\sum_{t}
\log p(y_t^d)
\end{align}
as the regularization term for the second stage of the training.
Although very simple, this regularization is expected to bring each element of $y^d$ corresponding to the same phoneme closer to each other.

To summarize, the loss function of the generator in StraGAN-VC+ASR $\mathcal{I}_G(G)$ is denoted as
\begin{align}
\label{star_gen2}
    \mathcal{I}_G (G) = &\mathcal{L}_{\text{adv}}^G (G) + \lambda_{\text{cls}} \mathcal{L}_{\text{cls}}^G (G) \nonumber\\
    &+ \lambda_{\text{cyc}} \mathcal{L}_{\text{cyc}} (G) + \lambda_{\text{id}} \mathcal{L}_{\text{id}} (G) \nonumber \\
    &+ \beta \mathcal{L}_{\text{ASR}} (G),
\end{align}
where $\beta\geq0$ is the hyperparameter that weighs the importance of the regularization term. 
In addition to the above, the loss functions for $D$ and $C$ are the same as those in Equations \ref{star_d_loss} and \ref{star_c_loss}, respectively.

\section{Experiments}
\subsection{Experimental setup}
\noindent \textbf{Dataset:}
We evaluated our method on a multi-speaker VC task using an ATR digital sound database~\cite{KUREMATSU1990357}, which consists of recordings of five female and five male Japanese speakers.
The spoken sentences were recorded as waveforms and were sampled at 20 kHz. 
We used a subset of speakers for training and evaluation.
We selected two female speakers, ``FKN'' and ``FTK,'' and two male speakers, ``MMY'' and ``MTK'', resulting in twelve different combinations of source and target speakers.
We selected 24 and 20 sentences for training and evaluation, respectively.
Therefore, there were $12\times 20 = 240$ test signals in total.

\noindent \textbf{Conversion process:}
36 Mel-cepstral coefficients (MCEPs), logarithmic fundamental frequency (log $F_{0}$), and aperiodicities were extracted every 5 ms using the WORLD analyzer~\cite{morise2016world} (D4C edition~\cite{morise2016d4c}).
In these experiments, we applied the baseline and proposed methods only to MCEP conversion. 
The $F_{0}$ contours were converted by logarithmic Gaussian normalization~\cite{liu2007high}. 
The aperiodicities were used directly without modification.
We trained the networks using the Adam optimizer~\cite{kingma2014adam} with a batch size of 8.
The number of iterations was set to $2\times 10^{3}$, the learning rates for $G$ and $D$ were set to $0.001$, and the momentum term was set to 0.5. 
We set $\lambda_{cls}=1.0$, $\lambda_{cyc}=1.0$, $\lambda_{id}=1.0$ following the original paper~\cite{kameoka2018stargan}.
We set $\beta=0.01$ empirically. 

\noindent \textbf{Implementation details:}
We used a dictation-kit (version 4.5)\footnote{https://osdn.net/dl/julius/dictation-kit-4.5.zip} as the trained ASR system. This package contains executables of Julius (deep neural network (DNN) version)~\cite{lee2009recent}, Japanese acoustic models (AM), and Japanese language models (LM).
Julius is a high-performance, small-footprint, large-vocabulary continuous speech recognition (LVCSR) decoder software used by speech-related researchers and developers.
The AM is a speaker-independent triphone DNN-hidden Markov model (H-MM) trained using the JNAS and CSJ corpora. 
It also has regression tree classes that are required for speaker adaptation by HTK\footnote{Hidden Markov Model Toolkit (HTK), http://htk.eng.cam.ac.uk/}. 
The LMs are 60k-word N-gram language models trained using the BCCWJ corpus.
Moreover, we set the dictionary data, which consisted of only utterances of training data, to improve the accuracy of speech recognition.

\begin{figure}[t]
  \begin{center}
    \includegraphics[width=0.9\linewidth]{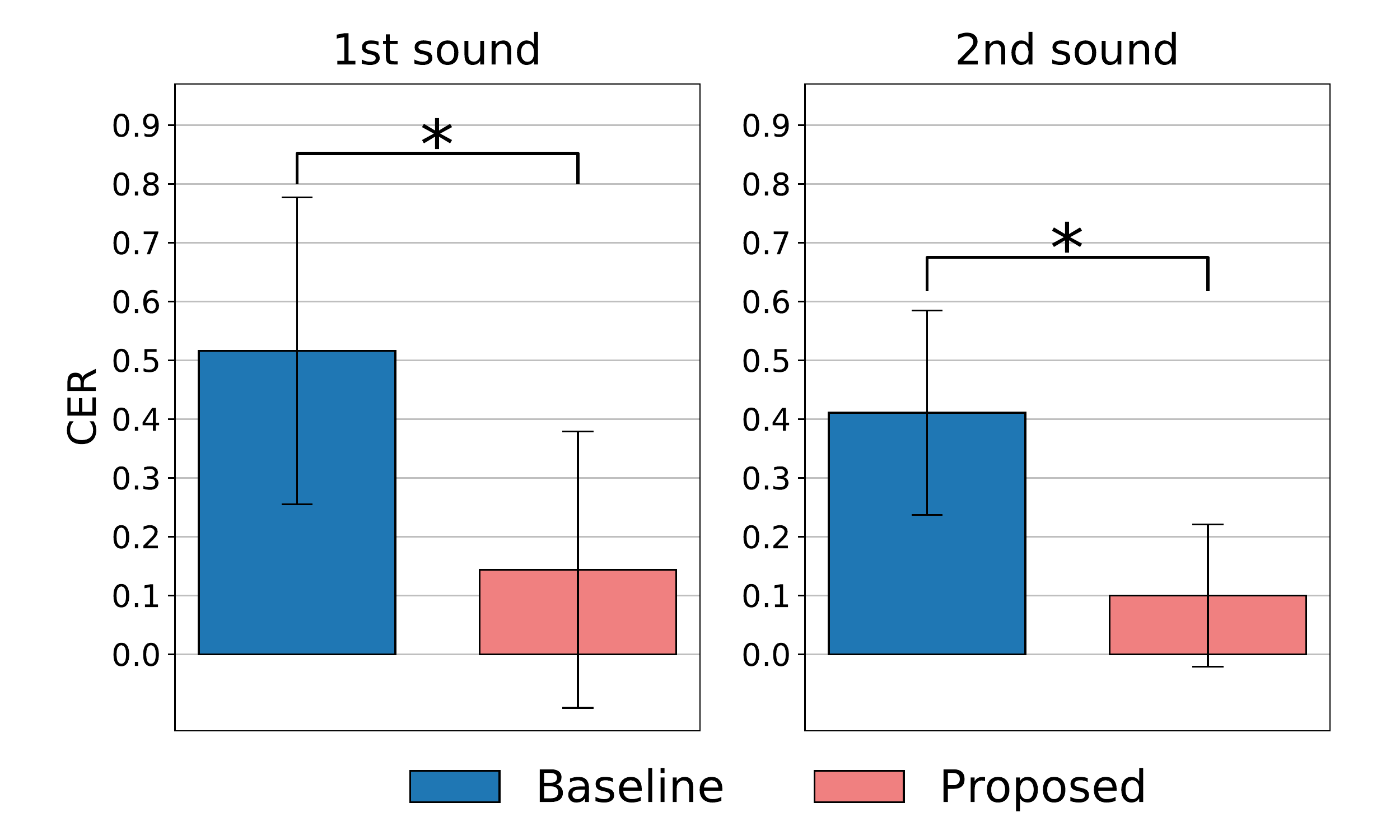}
    \vspace{-2mm}
    \caption{ Character error rates (CERs) for retention of linguistic information. CERs of the 1st and 2nd sound represent the first and second sounds heard by the listeners, respectively, and the content uttered are the same. The asterisks, *, indicate that the coefficients are statistically different from zero at the 1 percent level.}
    \label{cer_result}
  \end{center}
\end{figure}

\begin{figure}[t]
  \begin{center}
    \includegraphics[width=40mm, height=40mm]{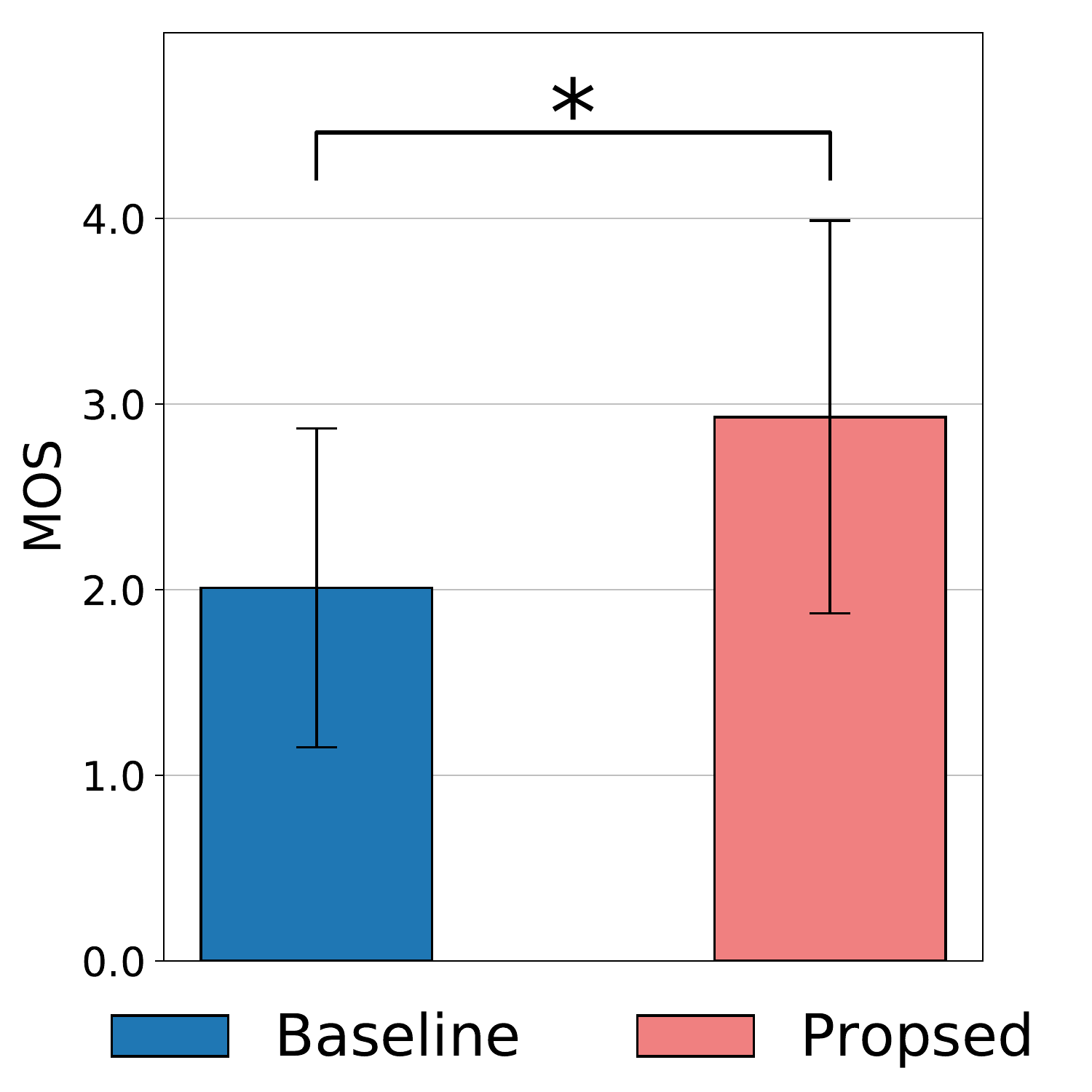}
    \vspace{-2mm}
    \caption{Result of the MOS test for naturalness. The asterisks, *, indicate that the coefficients are statistically different from zero at the 1 percent level.}
    \label{mos_result}
  \end{center}
\end{figure}

\subsection{Subjective evaluation}
We conducted listening tests to analyze the performance of the proposed method compared to StarGAN-VC~\cite{kameoka2018stargan}, which is the baseline in our proposed method. 
To measure the consistency of the linguistic contents of input and converted speech, we used the character error rate (CER) test as an evaluation metric. 
CER evaluates the error rate between the correct sentence and the sentence written by a listener after listening to the converted speech.
During the CER test, the listeners hear the same utterance of a converted speech twice because they may not be able to hear the converted speech at once.
The first trial is denoted as ``1st sound,'' and the second trial is denoted as ``2nd sound'' in Figure 4.
To measure naturalness, we conducted a mean opinion score (MOS) test.
During the MOS test, listeners were asked to rate the naturalness of the converted speech on a 5-point scale.
To measure the clarity of linguistic information, we conducted an AB test. 
``A'' and ``B'' were the speech converted by the baseline and proposed methods, respectively.
For each sentence pair, the listeners were asked to select their preferred one, ``A,'' ``B,'' or ``Fair.''
We conducted CER and MOS tests as in Experiment 1, and the AB test as in Experiment 2.
Ten sentences were randomly selected for Experiments 1 and 2 from 240 synthesized speech signals.
Sentences of different utterances were selected between Experiment 1 and Experiment 2.
Sixteen well-educated Japanese speakers participated in the tests. 

\noindent \textbf{Results of Experiment 1:}
Figure~\ref{cer_result} and Figure~\ref{mos_result} show
the average CER score for the retention of linguistic information and MOS for naturalness, respectively. 
As shown by the results of the CER test, the proposed method significantly outperformed the baseline method in terms of retaining the linguistic information of both 1st and 2nd sounds. 
In addition, the MOS test shows that the proposed method outperformed the baseline method in terms of naturalness.

\begin{figure}[t]
  \begin{center}
    \includegraphics[width=\linewidth]{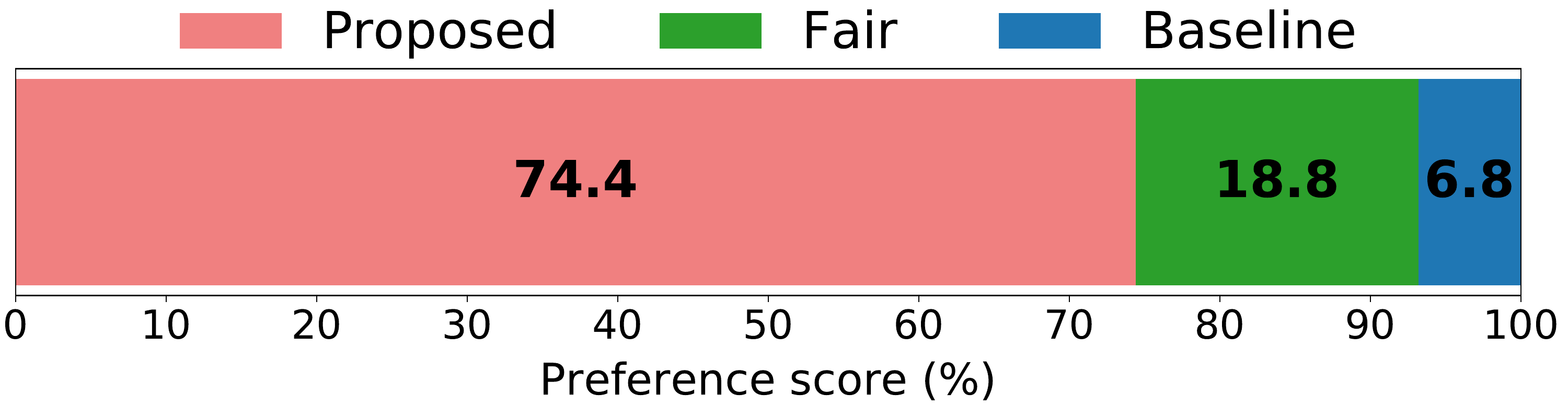}
    \vspace{-6mm}
    \caption{Results of the AB test for the retention of linguistic information}
    \label{exp2}
  \end{center}
\end{figure}

\noindent \textbf{Result of Experiment 2:}
Figure~\ref{exp2} shows the score of the AB test in terms of the clarity of linguistic information.
Based on the score, the proposed method achieved the highest score in terms of the clarity of linguistic information.

\section{Conclusion}
The original StarGAN-VC can fail to preserve the linguistic content of input speech when the number of training utterances is very small. To address this issue, in this paper, we proposed StarGAN-VC+ASR, a method that exploits the result of ASR applied to each training utterance during the training of StarGAN-VC's generator. 
Experimental results showed that StarGAN-VC+ASR had a better ability to preserve linguistic contents than the original StarGAN-VC.

In the present experiment, we used a small dataset for training and showed that our method outperformed StarGAN-VC. However, as the size of the dataset increases,  the performance of StarGAN-VC is also expected to improve.
In the future, we plan to investigate how the performance of StarGAN-VC and StarGAN-VC+ASR changes as the size of the dataset increases.


Moreover, to improve the quality of sounds converted by the proposed method, 
employing a mel-spectrogram and neural network vocoder~\cite{tamamori2017speaker,prenger2019waveglow,kumar2019melgan} instead of the MCEPs and WORLD analyzer, respectively, can be a future challenge.



In this study, we applied StarGAN-VC+ASR to speaker-identity voice conversion. However, the application of non-parallel voice conversion is not limited to speaker-identity voice conversion. For example, emotional voice conversion is an important target for voice conversion. The application of StarGAN-VC+ASR to emotional voice conversion will be our future work. StarGAN-VC+ASR used phoneme index information as a regularization term. This term improved the clarity of the converted voice. However, this may reduce the diversity of para-linguistic expressions in utterances and negatively affect emotional voice conversion. Validating this point will also be part of our future work.

\section{Acknowledgements}
This study was supported by the Japan Society for the Promotion of Science (JSPS) KAKENHI Grant-in-Aid for Scientific Research (B), grant number 18H03308, Grant-in-Aid for Scientific Research on Innovative Areas (grant number 16H06569), and JST CREST Grant JPMJCR19A3.


\bibliographystyle{IEEEtran}

\bibliography{main}


\end{document}